\documentclass[aps,pra,showpacs,superscriptaddress,preprint,showkeys]{revtex4}
\usepackage{graphicx}
\usepackage{dcolumn}
\usepackage{bm}
\begin{document}
\title{Economic thermodynamics and inflation}

\author{I. Demir\footnote{idris.demir@asbu.edu.tr}}
\affiliation{Department of International Relations, Ankara Sosyal
Bilimler University, Ankara, TR-06000, Turkey}

\author{A. I. Keskin\footnote{aliihsan.keskin@batman.edu.tr; alikeskin039@gmail.com (corresponding author)}}
\affiliation{Department of Electrical and Energy, Besiri OSB
Vocational School, Batman University, Batman, TR-72060, Turkey}

\begin{abstract}
This study presents a computational and theoretical framework
inspired by thermodynamic principles to analyze the dynamics of
economic inflation within adiabatic and non-adiabatic systems. In a
framework referred to as developmental symmetry, inflation is
formulated as a scalar field evolving through continuity equations,
drawing an analogy with the Raychaudhuri equation in gravitational
dynamics. The results show that adiabatic systems fail to reach
equilibrium, while non-adiabatic systems can evolve toward stable
states over time. The model successfully reproduces observed
inflationary regimes-from hyperinflation to stable low-inflation
phases-with characteristic transition periods of about a decade.
These results indicate that production continuity and controlled
monetary flow are crucial for achieving stability in complex
economic systems, linking thermodynamic balance to macroeconomic
equilibrium.
\end{abstract}


 \maketitle
\section{Introduction}
Inflation is characterized by a rapid decline in the purchasing
power of money and a seeming monotonous upward movement of prices.
Inflation has occurred frequently throughout history\cite{R1}, and
some nations are still dealing with its threat. So-called
hyper-inflation, in particular, is a type of monetary system failure
that is detrimental to society as a whole. Hyperinflation models are
particularly useful for highlighting how inflation indicates poor
''states of natural system'' in the economy. Cases of such a system
are characterized by wars, social regime shifts, state bankruptcies,
etc. As can be seen in the analyzed
situations\cite{R2,R3,R4,R5,R6,R7,R8,R9,R10}, these elements, along
with the influence of pure economic variables like expectations,
money demand, velocity of circulation, and quantity of money, result
in a rise in the consumer price index that is greater than
exponential. This behavior therefore has a detrimental impact on the
social network, creating awkward circumstances. Since an exact
scientific definition of inflation has not yet been established,
there is currently no method available to detect hyperinflation in
its early stages, which could help prevent the tragic event. The
literature-available model for hyperinflation is based on a
nonlinear feedback indicated by a power-law exponent with a value
greater than 0. In such a method, it can be mention that the
consumer price index displays a finite temporal singularity of the
type, $\frac{1}{(t_{c}-t)^{\frac{1-\alpha}{\alpha}}}$, enabling the
identification of a critical moment $t_{c}$ at which the economy
would collapse. Numerous situations where this model has been used
successfully can be observed in studies
\cite{R4,R5,R6,R7,R8,R9,R10}. Real economic activities are harmed
when inflation rises over modest levels. For example, because tax
revenues are effectively received over a longer period after the
announcement that fixes them, they have an impact on government
revenue\cite{R11,R12}. Since it is difficult to tell whether a price
increase is due to a relative price shift or is a result of general
inflation, perceiving relative price changes becomes more
challenging, which is called the Lucas problem\cite{R13}. If the
adjustment process varies for different types of goods, it results
in an inaccurate allocation of resources\cite{R14} and leads to
wasteful changes in relative prices\cite{R15}. A currency's
qualities such as a store of value, a medium of exchange, and a unit
of account are all impacted by inflation. As a result, the level of
disturbance increases as inflation
increases.\\
The thermodynamic analysis of an economic system (the first and
second law, thermodynamic equilibrium and processes) and analysis
with methods in statistical physics can be seen in the
literature\cite{R16,R17,R18,R19,R20,R21,R22,R23,R24,R25,R26,R27,R28,R29,R30,
R31}. In general, it can be said that the analysis of both economic
and social systems, with their appropriate definitions in physical
sciences, is a form of ''developmental symmetry''. In particular, by
drawing an analogy with its application to physics, it has been
demonstrated in work \cite{R30,R31} that the thermodynamics of
markets can be constructed as a phenomenological theory.\\
In this study, it is shown that Raychaudhuri equation
\cite{R32,R33,R34,R35} can describe an economic system without going
into deep mathematical proofs. The solutions of the equation for
adiabatic and non-adiabatic processes that occur in the energy
conservation equation, which are considered in the thermodynamic
framework, are investigated. At this point, it has been observed
that the adiabatic process in which hyperinflation emerged can be
eliminated by certain economic processes and monetary policies
directing the system. In this direction, we consider the economic
system as a field that can be defined as a scalar field $\phi$. The
existence of a minimum inflation in this field is a necessity even
at the beginning. The Raychaudhuri equation appears as an equation
describing such an economic system that can bend and expand
continuously. When we say bending or curvature of the field, we mean
whether the unit (point A in Figure 1) with kinetic energy (monetary
purchasing power) can reach a point with potential energy
(movable-real estate, valuables) within the field. In this context,
the present study aims to formulate an analytical framework in which
macroeconomic behavior can be interpreted through the geometrical
properties of the economic field. This approach provides a novel
perspective for understanding inflationary dynamics and market
evolution in terms of curvature and expansion processes. Thus, the
proposed analogy offers a new interpretive framework connecting
economic stability and inflationary behavior with field dynamics,
suggesting that economic equilibria may be viewed as geometrical
configurations within this scalar field. Therefore, this analogy
provides a foundation for modeling economic evolution as a geometric
process, allowing inflationary behavior and market expansion to be
understood within a unified thermodynamic and relativistic
framework.
\section{The Raychaudhuri Equation and Inflation}

In general, inflation is a concept used to explain the continuous
increase in the average price level within an economy. At the same
time, this increase implies a decline in the value of money (Reserve
Bank of New Zealand \cite{R36}). In the mathematical constructs of
economics, rather than relying solely on theoretical proofs,
conceptual economic phenomena are structured within the framework of
general system theory.

It should be noted that the present work does not aim to establish a
strict physical equivalence between cosmological and economic
variables. Instead, the analogy developed here is used
phenomenologically to describe the qualitative behavior of economic
expansion and contraction processes through the lens of dynamic
systems.

With these intuitive and reasonable assumptions, economic systems
are examined as evolving entities that can be analyzed using methods
inspired by physical sciences. We will mathematically restructure
the dynamics of inflation, defined as the persistent rise in the
prices of movable and immovable goods \cite{R31}, not as a process
limited to a subsystem, but as an intrinsic property of the general
structure of the economy, represented within a continuous field-like
framework.

Here, the economic variable $x(t)$ represents the relative distance
between purchasing power and the price level, analogous to a
system-wide growth parameter describing overall economic expansion.
This definition allows inflation to be expressed in terms of the
time evolution of this field.

In this framework, the purchasing power is represented by point $A$
in the field, and the price of goods by point $B$. The distance
between both points is scaled by $x$. The availability of $B$ in the
economic field (i.e., the presence of goods and services) is assumed
to prevent deflationary collapse. The system is considered to be in
an inflationary phase when the acceleration term satisfies
$\ddot{x}>0$, and in a deflationary phase when $\ddot{x}<0$. An
increase in the $x$ scale under the condition $\ddot{x}>0$ indicates
a decrease in purchasing power, thus characterizing inflationary
behavior.

Consequently, the acceleration of the economic field can be viewed
as a measure of inflationary pressure, where fiscal and monetary
interventions act as external forces capable of modulating the
field's expansion rate.

In the literature, the relative rate of change in the price of goods
is generally defined as
$$
C=\frac{\dot{p}}{p}
$$
\cite{R37,R38,R39,R40}. According to the present framework, an
alternative definition of the inflation rate can be expressed as
$$
C=\frac{\dot{x}}{x}
$$
which shows how an economic system or equivalently, an inflation
field evolves over time. In such a system, the establishment of a
developed economic equilibrium depends on the system's ability to
convert production value, represented by $B$, into purchasing power
at point $A$.

In an inflationary context, the conversion of movable or immovable
assets (including technological goods) into monetary resources often
occurs through export activities or external resource inflows. These
represent non-adiabatic processes within the global economic
environment. On the other hand, it can be observed that the
accessibility of $A$ to $B$ decreases continuously, since $C$ tends
to decline over time under inflationary conditions. This behavior
can, for instance, be modeled as
$$
x\simeq t^{h}
$$
where a decrease in $C$ occurs at a rate
$$
\frac{h}{t}
$$
when $h>1$. Therefore, the economic system exhibits a dynamic
structure characterized by the scaling behavior of $x$.

\begin{figure}[ht]
\centering
\resizebox{0.6\textwidth}{!}{%
\includegraphics{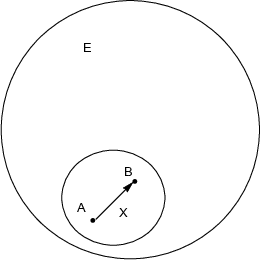}}
\caption{Conceptual representation of a subsystem within the
economic system $E$. Points $A$ and $B$ indicate two representative
positions in the system, scaled by $x$. The element at $A$
corresponds to purchasing power, while $B$ represents the value of
goods or assets. During inflationary expansion, systemic forces act
outward, reflecting a general increase in the scale of economic
transactions. Fiscal and monetary interventions function as
regulatory mechanisms that help maintain balance between purchasing
power and asset prices within this dynamic structure.}
\end{figure}

The energy density and pressure of this homogeneous and dynamic
economic field are given as follows \cite{R41,R42},
$$
\rho=\frac{\dot{\varphi}^{2}}{2}+V(\varphi),\,\,\,\,\,\,\,\,\,\,p=\frac{\dot{\varphi}^{2}}{2}-V(\varphi)
$$
Here, the kinetic term $\frac{\dot{\varphi}^{2}}{2}$ corresponds to
the purchasing power at point $A$, while the potential energy
$V(\varphi)$ represents the accumulated value of movable and
immovable goods at point $B$. A decrease in $C$ reflects reduced
accessibility between purchasing power and goods prices. Inflation
occurs when
$$
\frac{\dot{\varphi}^{2}}{2}<V(\varphi)
$$
whereas an economic equilibrium is achieved when
$$
\frac{\dot{\varphi}^{2}}{2}\simeq V(\varphi)
$$
The former corresponds to a phase of rapid expansion in the
$x$-scale, characterized by high prices and low purchasing power,
while the latter represents a stable, slowly expanding economy
typical of developed systems. Therefore, the first case produces
negative pressure associated with inflation (field expansion),
whereas the second results in positive pressure reflecting
purchasing power and economic stabilization.

A state parameter can be defined as
$$
\gamma=\frac{p_{\varphi}}{\rho_{\varphi}}
$$
which denotes the general state of the system. When $\gamma<0$, the
system is in an inflationary (expanding) phase, while a positive
$\gamma$ indicates a non-inflationary or decelerating condition.

Therefore, the analytical framework developed here allows inflation
to be interpreted as a curvature-like expansion within an abstract
economic field. This analogy highlights the role of fiscal and
monetary interventions as control parameters that regulate the
dynamic balance between purchasing power and asset values. While the
proposed model remains phenomenological, it provides a conceptual
bridge connecting macroeconomic behavior with the systemic dynamics
of complex evolving structures observed across both physical and
social domains.\\
We start with the Raychaudhuri equation\cite{R32,R33,R34,R35}, which
describes the motion of localized units (densities) at points $A$
and $B$ defined in the field, and describes the motion between two
points. For a homogeneous distributed dynamic system, this equation
can be written as follows:

\begin{equation}
\frac{d\theta}{d\tau}=-\frac{1}{3}\theta^{2}-4\pi(\rho_{\varphi}+3p_{\varphi}).
\end{equation}
where $\theta=\frac{\dot{x}}{x}$. Therefore, it can be written the
equation,

\begin{equation}
\frac{-3\ddot{x}}{x}=4\pi(\rho_{\varphi}+3p_{\varphi}).
\end{equation}
However, to achieve the continuity equation of the perfect fluid we
make use of the first law of thermodynamics,
\begin{equation}
\delta Q=dE+pdV=TdS,
\end{equation}
where $E$ is internal energy of subsystem (amount of money in the
system), $V$ is the volume of system (the amount of goods in
system), $p$ is the pressure of system (the indicator of purchasing
power in the system) and $\delta Q$ is heat follow (money transfer
that occurred  between the subsystems or systems without buying and
selling goods). Hence, since a macro-level observer will observe the
energy from these quantities as a density dispersed in the volume in
the subsystem, under the approximation $d E\simeq dm\simeq
Vd\rho+\rho dV$, the first law is read as follows,
\begin{equation}
\dot{\rho}+3C(\rho_{\varphi}+p_{\varphi})=\frac{\delta Q}{Vdt},
\end{equation}
This is the conservation equation of the economic system, which
states that the quantities of the system are conserved during the
economic cycle with time. Two solutions of equation (5) can be
investigated for adiabatic ($\delta Q\simeq 0$) and non-adiabatic
($\delta Q\neq 0$) processes. The adiabatic case where there is no
money transfer at the borders of the system means that there is no
money transfer economically to a country without buying or selling
goods. The other non-adiabatic process means there is a flow of
money to or from country borders. In the next two sections, we will
discuss the solutions of these two processes from an economic point
of view.
\subsection{Adiabatic process}
The continuity equation for the adiabatic process takes the
following form,
\begin{equation}
\dot{\rho}+3C(\rho_{\varphi}+p_{\varphi})=0,
\end{equation}
and whose solution is as follows
\begin{equation}
\rho_{\varphi}=\rho_{0}a^{-3(1+\gamma)}.
\end{equation}
Here, $\rho_{0}$ is an integral constant. Then, the eq. (5) takes
the form,
\begin{equation}
-3\ddot{x}= 4\pi \rho_{0}x^{-3(1+\gamma)+1}(1+3\gamma),
\end{equation}
So, we proceed by finding some solutions for the eq. (9) to
investigate inflation. Using the modulating parameter,$w$, we can
write $\dot{\varphi}^{2}(1-\gamma)=2(1+\gamma)V(\varphi)$. This
equation tells us that there exists a boundary in an economic system
at the point $\gamma\neq 1$ because the case $V(\varphi)=0$
representing movable or immovable property is not considered in an
economic system. But, it is possible in the case $\gamma\sim -1$,
which corresponds to a hyperinflation. This is a case where the
monetary purchasing power represented by kinetic energy,
$\dot{\varphi}^{2}$, is almost nonexistent or capital becomes almost
worthless. Therefore, the solutions close to $\gamma\sim -1$ produce
exponential hyperinflation solutions,

\begin{equation}
x\simeq \exp(\sqrt{\frac{8\pi \rho_{0}}{3}}t
)C_{1}+\exp(-\sqrt{\frac{8\pi \rho_{0}}{3}}t )C_{2},
\end{equation}
where $C_{1}$ and $C_{2}$ are integral constants. The first
solution, which shows the increase in the distance scale between the
purchasing power and the price of the good, is considered. In other
words, an increase in scale $x$ means that there is a decreasing
effect on purchasing power due to $\gamma\sim -1$
($\frac{\dot{\varphi}^{2}}{ 2 }\ll V(\varphi)$). Therefore, we can
set $C_{2}$ to zero, where the solution $x\simeq
\exp(\sqrt{\frac{8\pi \rho_{0}}{3}}t )C_{1}$ survives. This appears
in Figure 1 with $C_{1}=1$.
\begin{figure}[ht]
\centering
\resizebox{0.6\textwidth}{!}{%
\includegraphics{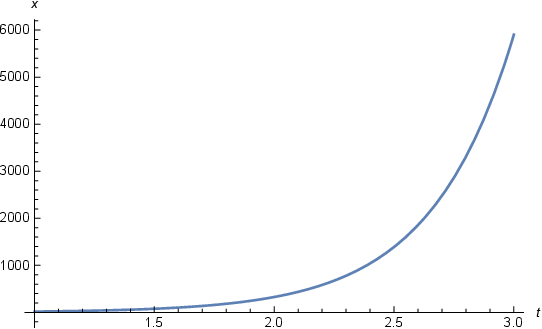}}
\caption{The relationship between the scale variable $x$ and time
$t$. The exponential increase in $x$ represents the
hyperinflationary phase of the system. In this regime, purchasing
power, interpreted as the system's kinetic component, becomes almost
negligible, while the rapid increase in prices is driven by
accumulated asset values and capital concentration, corresponding to
the potential component of the system.}
\end{figure}
The high potential indicates high prices of material or financial
resources that have not yet been converted into monetary resources,
that is, into the system's kinetic component. An exponential
increase in the scale variable $x$ indicates a loss of purchasing
power. This unstable situation can be mitigated by reducing demand
for goods, which in turn leads to a decline in potential, that is, a
decrease in prices.

In an adiabatic system meaning no external monetary inflow or
outflow the potential (price level) can be reduced either by
increasing production or by lowering demand. It should be emphasized
that artificially forcing the system into a non-adiabatic state is
not an effective solution. Although a temporary monetary inflow (for
example, capital entering the system through interest-based
borrowing) can increase the kinetic component (purchasing power) in
the short term, it ultimately leads to stronger inflation in the
long run. This is reflected in Figure 2, where the curve peaks over
time.

Furthermore, temporary external monetary inputs distort the
homogeneity of the system by introducing artificial kinetic energy,
which eventually drains resources through interest payments and
creates structural imbalance. Monetary inflows enrich one segment
while exploiting another, leading to an inhomogeneous field where
purchasing power continues to erode. Hence, sustained inflation
persists, especially for units whose kinetic energy (purchasing
capacity) declines more rapidly.

Under adiabatic conditions, the most realistic path toward restoring
equilibrium involves reducing demand encouraging economic savings
and consumption restraint. This drives sellers to lower prices to
convert their assets into money, which is a key mechanism for
reducing inflation. In addition, increasing the availability of
natural or productive resources decreases potential (prices) and
enhances monetary valuation, moving the system toward stability.

However, wasting or underutilizing resources contradicts the policy
direction of minimizing consumption. Therefore, satisfaction with
existing goods either voluntarily or as a result of policy
pressure becomes a natural adjustment mechanism.

The rate of inflation in this regime is given by $C = \rho_{0}$. If
$\rho_{0}$ is fixed at a large value, inflation remains nearly
constant over time. Thus, under rational policy frameworks, the rate
of change $C$ should take a time-dependent form \cite{R42} to allow
for gradual adjustment. In developed economies, the system tends to
reach an equilibrium where purchasing power equals potential,
$\frac{\dot{\varphi}^{2}}{2} \simeq V(\varphi)$, corresponding to
$\gamma \simeq 0$. However, the solution of Eq. (10) near this point
($\gamma \simeq 0$) is an empty set.

Therefore, although $K = V$ holds approximately in adiabatic
systems, under rational fiscal policies--such as demand reduction
through savings--the system evolves toward a linear behavior at
$\gamma \simeq -\frac{1}{3}$, where pressure gradually falls:
\begin{equation}
x\sim C_{1}+tC_{2},
\end{equation}
Taking the real part ($C_{1} = 0$), we obtain $x \sim tC_{2}$. The
ratio parameter takes the form $C = \frac{1}{t}$, showing a
decreasing trend over time. As a result, fiscal restraint policies
that are maintained over long periods can effectively lower
inflation.

If $t$ is measured in years, this model suggests that inflation
could decline to single-digit levels over a period longer than ten
years. As will be discussed in the following section, when moderate
inter-system trade is allowed, an economic equilibrium point may be
reached at $K = V$, and inflation may decline more rapidly within
less than a decade even without strict fiscal constraints.

\subsection{Non-adiabatic process}

We shall use the entropic force definition to reveal how the term on
the right-hand side of eq. (7). The entropic force is defined as
follows,

\begin{equation}
F_{ent.} = -\frac{T dS}{dx},
\end{equation}

where for a non-adiabatic process we have $T dS = \delta Q \neq 0$,
so the eq. (7) can be written as follows,

\begin{equation}
\frac{d\rho}{\rho} + 3(1+\gamma) \frac{dx}{x} = - \frac{F_{ent.}
dx}{E}.
\end{equation}

Herein, due to $E \simeq \rho V \simeq F_{ent.} x$, where recall
that $\rho = kinetic + potential$, we can write

\begin{equation}
\frac{d\rho}{\rho} + 3(1+\gamma) \frac{dx}{x} = - \frac{dx}{x}.
\end{equation}

The solution of this equation is as follows,

\begin{equation}
\rho = \rho_0 x^{-3(1+\gamma)-1},
\end{equation}

with an integral constant $\rho_0$. Thus, the eq. (5) takes the
form,

\begin{equation}
-3 \ddot{x} = 4 \pi \rho_0 (1 + 3\gamma) x^{-3(1+\gamma)},
\end{equation}

Hyperinflation solution $\gamma \simeq -1$ is

\begin{equation}
x = \frac{4}{3} \pi \rho_0 t^{2} + C_1 + C_2 t,
\end{equation}

where we set $C_1 = C_2 = 0$, so an increase in $x$ is parabolic,
indicating very low economic purchasing power.

\begin{figure}[ht]
\centering
\resizebox{0.6\textwidth}{!}{%
\includegraphics{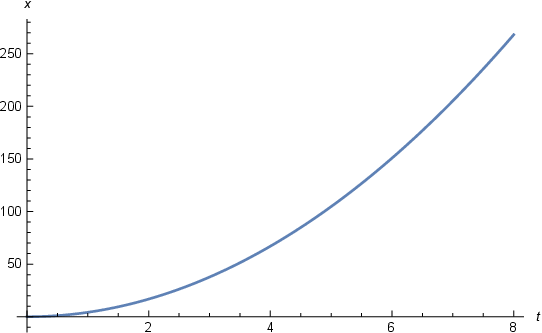}}
\caption{Figure showing the increase of the x scale in the (0-8)
time interval.}
\end{figure}

The proportional parameter of this type of inflationary field
behaves as $C = \frac{2}{t}$. Even if the inflation parameter $C$,
which starts in a zero time instance, returns to its natural
appearance after 20 years, it reveals a deep deformation situation
that ruins the field due to the size of the $x$ scale. Therefore, in
such an inflationary field, it is necessary to intervene in the
system with reasonable fiscal policies and to pass the inflationary
phase to a new phase. In this way, a stable course can be followed
by keeping the economy under control without any deformation.

Therefore, when we apply the $K=V$ or $w \simeq 0$ approximation
that we discussed for economic equilibrium to the equation, we get
the following equation.

\begin{equation}
x = \pm \left( \frac{-4 \pi \rho_0 + 3 t^{2} C_1^{2} C_2 + 6 t
C_1^{2} C_2 + 3 C_1^{2} C_2^{2}}{3 C_1} \right)^{\frac{1}{2}}.
\end{equation}

Note that the solutions (17) and (18) converge at about $t=4$. We
can manipulate the value of $C_1$ based on any value of $C_2$. For
example, the value graph at $C_1=0.4$ for $C_2=1$ in the wide
selected range of $x=(4-15)$ is given in Figure 1. If $C_2=2$ is
chosen, $C_1=0.34$ produces the same graph. This means that
inflation entered a declining phase (from $t \sim 4.1$ to about 11)
and progressed in a developed economy phase after $t \sim 11$.

\begin{figure}[ht]
\centering
\resizebox{0.6\textwidth}{!}{%
\includegraphics{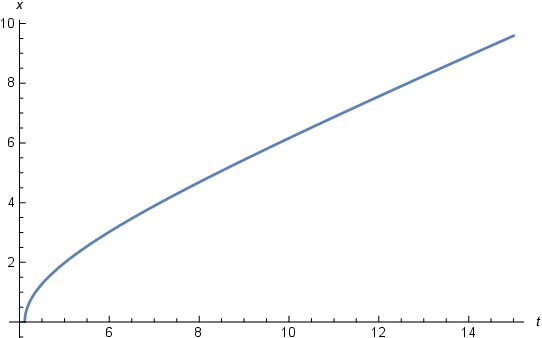}}
\caption{The figure shows the behavior of the x scale in the time
interval (4-15). Here, $C_1=0.4$ is chosen.}
\end{figure}

We can again read the change in the inflation rate from the
parameter $C = \frac{\dot{x}}{x}$. If we consider $t$ in terms of
years, the economy with the traces of hyperinflation between $t=4.1$
and $t=5$, and a rate that decreases to 10 percent in approximately
7 years between $t=4.1$ and $t \simeq 11$ is obtained. After the
approximate value of $t \simeq 11$, a single-digit inflation value
may appear. This indicates the economic equilibrium represented by
$K=V$ for the economic system. Let's note that it takes about 10
years for it to reach equilibrium from a state of almost no
purchasing power during the hyperinflation period and that it takes
place with the external heat transfer to the system (occurring money
transfer without buying-selling cases). Let's remember that the
adiabatic process did not produce the $K=V$ solution. In such
processes, we emphasized that inflation can be avoided by reducing
the demand for purchasing goods.

\begin{table}[h!]
\centering
\begin{tabular}{l l}
 \hline
 Country & Inflation period \\
 \hline\hline
 Bolivia     & 1970--1985 \\
 Peru        & 1970--1990 \\
 Israel      & 1970--1985 \\
 Brazil      & 1970--1994 \\
 Nicaragua   & 1985--1991 \\
 Hungary     & 1946 Jan--1946 Jul \\
 Germany     & 1922 Nov--1923 Oct \\[1ex]
 \hline
\end{tabular}
\caption{Inflation periods in certain year intervals of some
countries \cite{R3}.} \label{table:1}
\end{table}

With the decrease in demand for the density units (energy, i.e.,
money) that describe the system, the kinetic energy or purchasing
power will increase over time. Another possible scenario is
political intervention aimed at reducing demand for the purchase of
goods. Such policies reduce purchasing demand gradually, leading
consumers to limit their needs, which increases kinetic energy over
time and thus enhances purchasing power.

However, in the non-adiabatic process, external monetary investments
or the presence of rich resources (potential assets convertible into
money, i.e., kinetic energy) within the system will steer the
economy toward balance over time, as our solutions have
demonstrated.

Table~\ref{table:1} shows some countries that experienced
hyperinflationary periods. Considering these intervals, it is
observed that inflation typically persists for about 10 years,
consistent with our theoretical calculations.

For example, under reasonable financial policies, the economic
system in Turkey, which entered a hyperinflationary phase starting in
2017, may be expected to complete its natural course after 7 years
and reduce inflation to single digits after approximately 11 years
(around 2028).

\section{Closing remarks}

There exist many symmetries in the universe (e.g. spatial and
temporal translation symmetries, parity, and charge symmetries,
etc.). Especially in physical science, symetry describes that when
any observable quantity is considered, some properties of the system
do not change under certain transformations (for example, Noether
symmetries). In other words, symmetry tells us that the internal
property of a physical system related to physics or mathematics does
not change with the change of some factors. In social science, which
is also a part of the universe, evaluating the welfare of society,
the interaction of the units that make up the social system
(including political interactions) and its development in terms of
its compatibility with the laws of theoretical physics, can offer a
realistic perspective. On the other hand, the physical development
of a plant or human, the growth of the circle of ideas, or the
expansion of the intellectual sphere of influence (just as the
expansion of the universe) refer to observable symmetries. In this
study, which we handled mathematically according to this point of
view, which we call "developmental symmetry" in the universe, we
discuss inflation as an ever-expanding system (similar to the
dynamic structure in developed economies) with its scalar field
representation. In this direction, we investigate the economics of a
system from an inflationary perspective within the framework of the
Raychaudhuri equation. We have obtained a continuity equation based
on the first law of thermodynamics. For the adiabatic and
non-adiabatic processes of this equation, we obtain some solutions
corresponding to the inflationary case for the system. As can be
seen in Figure 1, the main point we consider in the study is the
purchasing power, which is described by point $A$, and the purchased
goods, which is identified by point $B$. It should be noted that in
a developed economic system under consideration, the $x$-scale can
never reach a value of zero. Finding an $x$ value in developed
economies is important for the continuity and development of the
economic system. Therefore, it is inevitable for a dynamic system
(developed economic system) to have a constant value
($\frac{\dot{\phi}^{2}}{2}\simeq V(\phi)$) of purchasing power and
price of goods for an exchange of approximately $x$ value to take
place. An increase in the $x$ scale means that the availability of
the good decreases, that is, the purchasing power
($\frac{\dot{\phi}^{2}}{2}$) decreases or the price of the good
$V(\phi)$ is high. We represented this situation with
$\frac{\dot{\phi}^{2}}{2}\ll V(\phi)$, therefore, solutions in which
hyperinflation occurs in the case of $\frac{\dot{\phi}^{2}}{2}\ll
V(\phi)$. As emphasized, the equilibrium point is the case of
$\frac{\dot{\phi}^{2}}{2}\simeq V(\phi)$, which corresponds to a
deflationary area that reduces the rate of hyperinflation. Since
there is no energy (monetary) flow with external systems for an
adiabatic economic system, it has been revealed that it can get rid
of inflation in more than 10 years after it evolves into an
inflation area of $\gamma=-\frac{1}{3}$ with reasonable financial
policies within the system. Also considering the duration of
reasonable policies, it will probably take more than 20 years to get
out of the inflationary field. For the more realistic non-adiabatic
process (the system with energy-money flow), we obtained
mathematical results that are quite compatible with the inflation
processes seen in Table 1. This analogy between thermodynamic
balance and economic equilibrium also reflects the collective
behavioral tendencies of economic agents--such as consumption habits,
trust in fiscal policy, and market expectations--that determine how
energy (money) circulates within a society.\\
 On the other hand, the main reason why
some countries in Table 1 left the hyperinflation area in less than
11 years can be shown as the inflow of money into the country. For
example, the cash flow from the international monetary fund to the
country arising from production, natural resources, technological
infrastructure and the sound economic policies created as a result
can be shown. The reasons for the inflation that continues for more
than 11 years on average may be wrong fiscal policies and management
of the system (country) as an adiabatic system. In our current
study, the 11-year average period of the inflation process is
obtained within the framework of "developmental symmetry" for
non-adiabatic systems. After an average of 11 years of
hyperinflation, the economic system is evolving into a stable state
(single-digit percentages). As we have emphasized, one of the most
effective ways to stop hyperinflation in such an economic system
created identically (symmetrically) to mathematical tools is to
ensure production in the system. Whether such a system is adiabatic
or not (whether there is energy-money flow or not) may no longer be
important. The use of natural resources with the right policies or
the production of technological parameters that may meet the needs
can lead to a reasonable inflation rate without the flow of money
with interest to the inflation area where purchasing power drops to
almost zero. In another separate study, it can be discussed whether
inflation can be cut or not by finding the creation of the scalar
field density in the field. Because density is a quantity that
includes kinetics and potential due to $E\sim \rho$. Units of
density created in the field by the political ground controlling the
field represents the unity of purchasing power (kinetic energy,
point $A$ in Figure 1) and movable immovable property (potential
energy, point $B$ in Figure 1). Therefore, it is clear that this
unity emerging everywhere in the field will have a decreasing effect
on inflation. This highlights that maintaining the balance between
purchasing power and asset value throughout the economic field is
crucial for mitigating
hyperinflation and ensuring sustainable economic growth.\\

\section{Data Availability Statement} No Data associated in the manuscript

\end{document}